\documentclass[runningheads]{svmult}

\usepackage{graphicx}  
\usepackage{multicol}  
\usepackage{multind}
\usepackage{physprbb}  
\makeindex{s}             
\makeindex{o}

\begin{document}
\title*{Spiral waves in accretion discs - observations}
\toctitle{Spiral waves in accretion discs}
%
%
\titlerunning{Spiral waves in discs}
%
\author{Danny Steeghs}
\authorrunning{Danny Steeghs}
%
%
\institute{Physics \& Astronomy, Southampton University, Southampton SO17 1BJ, UK}

\maketitle              

\begin{abstract}

I review the observational     evidence for spiral structure  in   the
accretion    discs of  \index{s}{cataclysmic variable}cataclysmic
variables (CVs).  \index{s}{Doppler tomography}Doppler tomography  is
ideally suited to resolve and map such co-rotating patterns and allows
a straightforward comparison  with theory.  The  dwarf nova  IP Pegasi
presents  the best studied case,  carrying  two \index{s}{spiral arms}spiral arms  in a  wide  range of emission lines  throughout  its
\index{s}{outburst}outbursts.  Both arms appear at the locations where
tidally driven \index{s}{spiral arms}spiral  waves are  expected, with
the arm closest to the gas stream weaker in the  lines compared to the
arm  closest   to   the      companion.  Eclipse   data      indicates
sub-\index{s}{Keplerian orbit}Keplerian velocities  in the outer disc.
The    dramatic    disc structure  changes   in  \index{s}{cataclysmic
variable!dwarf nova}dwarf   novae   on timescales of   days  to weeks,
provide  unique      opportunities     for   our     understanding  of
\index{s}{angular momentum}angular momentum  transport and the role of
density waves on the structure of accretion discs.
I  present  an  extension  to  the Doppler tomography  technique that
relaxes one of the basic assumptions of tomography, and is able to map
modulated emission   sources.   This   extension allows   us    to fit
\index{s}{anisotropic radiation}anisotropic emission from, for example, \index{s}{spiral arms}spiral shocks, the \index{s}{irradiation}irradiated
companion star and disc-stream interaction sites.

\end{abstract}

\section{Accretion discs and \index{s}{angular momentum}angular momentum}

The energetic phenomena associated    with a wide range  of  accreting
systems  rely on the efficient   conversion of  potential energy  into
radiation and heat. In close binaries,  the deep potential well of the
compact object leads to mass transfer and accretion once the companion
star\index{s}{secondary star}  evolves  and \index{s}{Roche equipotential}Roche lobe overflow  commences.   The efficiency of
accretion is proportional to  the compactness, $M/R$, of the accreting
compact star with mass $M$ and radius $R$.
As matter spills over near the first \index{s}{Lagrangian point}Lagrangian point,  it sets off on
a ballistic  trajectory towards the  accretor. Its potential energy is
converted into kinetic energy,  but its net   \index{s}{angular momentum}angular momentum, due  to
orbital motion of  the mass donor\index{s}{secondary star},  prevents a  straightforward path to
the accretor.   The natural   orbit  for such   matter is  a  circular
\index{s}{Keplerian orbit}Keplerian orbit  corresponding to its   specific \index{s}{angular momentum}angular momentum.  Instead of  dumping material directly onto the  compact star, the primary \index{s}{Roche equipotential}Roche lobe is
slowly filled with  a near \index{s}{Keplerian orbit}Keplerian  disc.  Angular momentum needs to
be dispersed  within this  accretion disc  in  order  to allow  gas to
spiral inwards towards the compact star \cite{fkr}.

It is  the   detailed  process  of  \index{s}{angular momentum}angular momentum   transport  that
determines the structure of this accretion disc and therefore the rate
at which gas,  supplied from the mass  donor\index{s}{secondary star}, is actually  accreted by
the   compact   object.  Although  so fundamental   to  the process of
accretion   through discs,   our  understanding of    \index{s}{angular momentum}angular momentum
dispersal  is   very limited.   We  can  roughly  divide the  possible
physical  mechanisms that  may provide  the required angular  momentum
transport  into  two classes.  Those that  work  on  a local scale and
exchange \index{s}{angular momentum}angular momentum among neighbouring  parcels in the disc, and
those  that  rely on global, large    scale structures in  the disc.  The local
processes are commonly referred to  as `viscous processes' even though
it   was clear that the  molecular   \index{s}{viscosity}viscosity of  the accretion  disc
material itself was many orders of magnitudes too small \cite{livio94}.
Viscous interaction in the sheared \index{s}{Keplerian orbit}Keplerian disc allows some material
to  spiral inwards, losing  \index{s}{angular momentum}angular momentum, while excess momentum is
carried   outwards by   other parts  of    the  flow.  In the   famous
$\alpha$-parameterisation of  Shakura  \&  Sunyaev  \cite{ss73},  this
\index{s}{viscosity}viscosity was replaced   by  a single  dimensionless  constant,  which
allows one to solve the structure equations for thin, viscously heated
accretion discs\index{s}{accretion disc!thickness} \cite{pringle81}.

A  very  different way  of  transporting the   \index{s}{angular momentum}angular momentum is via
density waves in the disc.  In self-gravitating  discs, the ability of
density  waves to transport angular  momentum  is a direct result from
purely gravitational interaction between  the wave and the  surrounding
disc material  \cite{toomre69,bt87}.   Waves  of this type  can
still  transport   \index{s}{angular momentum}angular  momentum  in  the absence  of self gravity,
provided  some mechanism exists   that exchanges momentum between  the
wave  and the fluid. Sawada  et  al. \cite{sawada} conducted numerical
simulations of mass transfer via \index{s}{Roche equipotential}Roche lobe overflow  of inviscid, non self gravitating discs, and witnessed the
development of strong \index{s}{spiral arms}spiral shocks in the disc which were responsible
for  the bulk   of the   angular  momentum  transport throughout the flow. Such trailing \index{s}{spiral arms}spiral patterns are the  natural result of a tidal
deformation of the disc that is  sheared into a  \index{s}{spiral arms}spiral pattern by the
(near) \index{s}{Keplerian orbit}Keplerian rotation profile of the disc material \cite{savonije}.

In this review I aim to give an  overview of the observational efforts
to   study  such \index{s}{spiral arms}spiral    structures   in  the  discs   of  \index{s}{cataclysmic variable}cataclysmic variables (CVs).  Tomography is  ideally   suited   for  the study of    disc
structure, and the detection  of a \index{s}{spiral arms}spiral structure  in the disc of IP
Pegasi\index{o}{IP Peg} \cite{shh97}, 13 years after the work by Sawada et al., triggered
a renewed  interest in such models. Boffin,  this volume, will focus
on the theoretical  side of the issue  and the comparison  between the
observations and numerical simulations of \index{s}{spiral arms}spiral waves in discs.

\section{Prospect for detecting spiral waves in discs}

\begin{figure}
\begin{center}
\includegraphics[width=.66\textwidth]{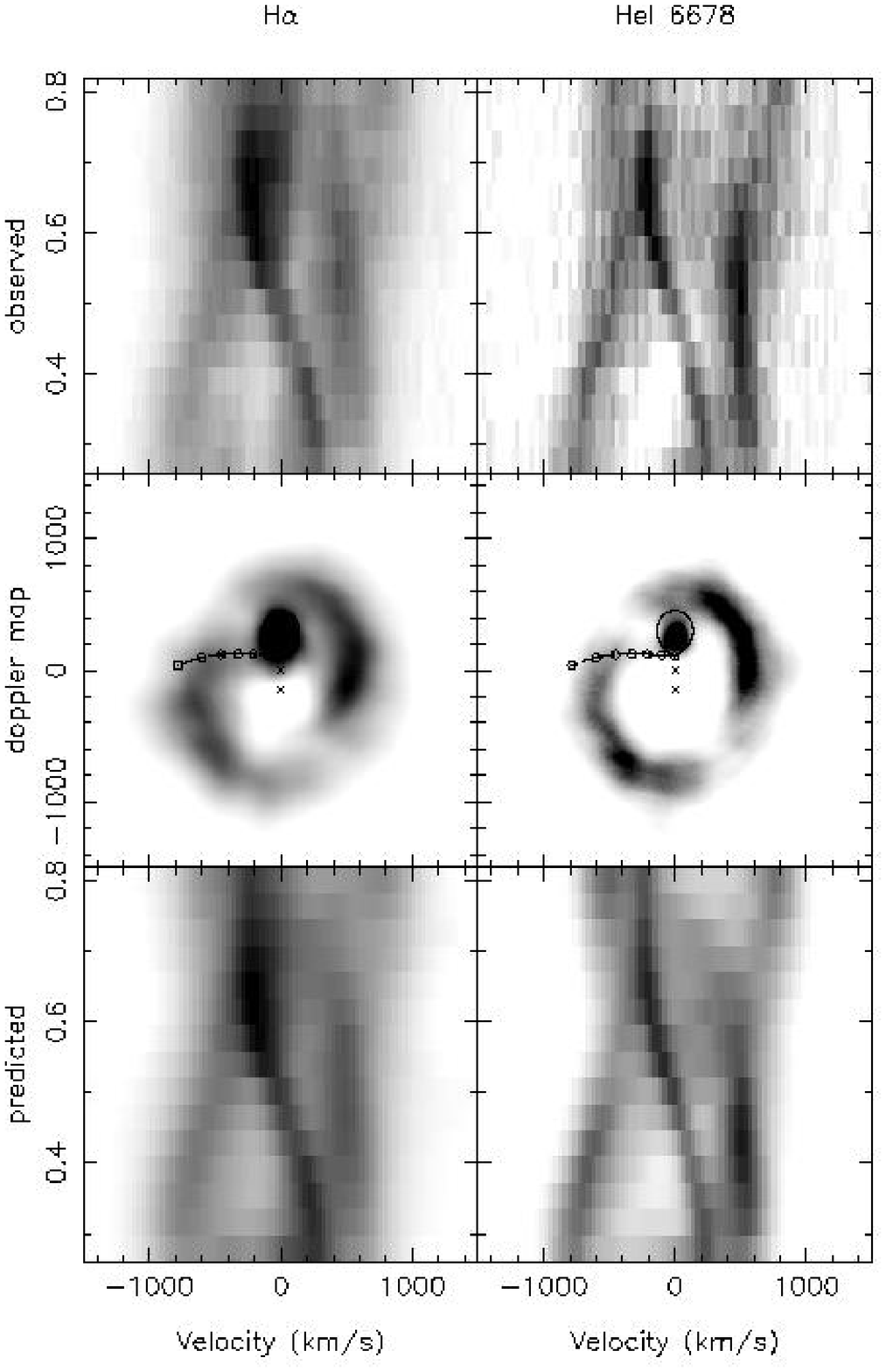}
\includegraphics[height=.6\textwidth,angle=-90]{figure1b.ps}
\end{center}
\caption[]{Top; the observed H$\alpha$ and HeI emission line profiles during the early stages of an  IP Pegasi\index{o}{IP Peg} \index{s}{outburst}outburst. The maximum \index{s}{entropy}entropy Doppler \index{s}{Doppler map}tomograms (second row) reveal a prominent \index{s}{spiral arms}spiral pattern in the disc as well as \index{s}{secondary star}secondary star emission. Third row of panels are the predicted line profiles given the calculated maps. Bottom row is a contour plot of the Doppler \index{s}{Doppler map}tomograms. From \cite{shh97}.}
\label{ip93}
\end{figure}

Tidally generated \index{s}{spiral arms}spiral waves are the result  of tidal torques of the
companion star on  the orbiting disc  material. Initially triggered at
the outer edge  of the disc, where  the tidal interaction between  the
disc and   the companion star  is   strongest, they take  the  form of
trailing  \index{s}{spiral arms}spirals because the azimuthal  velocity of the disc material
is  \index{s}{supersonic}supersonic and   increases monotonically  with decreasing \index{s}{distance}distance
from the   compact   object ($v_{\phi}\propto  r^{-1/2}$).   Both  the
density  as   well as the   disc  temperature are much  higher  at the
location of  the \index{s}{spiral arms}spirals, and  continuum  and line  emission from such
\index{s}{spiral arms}spirals can  thus  be expected   to be in    clear contrast  with  the
surrounding areas  of the disc.  This  contrast depends on the density
contrast in the  wave, i.e.  the  strength  of the  shock, as  well as
local radiative transfer conditions.

Although predicted   in  the 80s, observational evidence   for \index{s}{spiral arms}spirals
relies on  the  ability to  spatially resolve  the accretion  discs in
interacting binaries.  Indirect imaging methods   are thus required to
search for such global disc asymmetries.   Since the wave pattern is a
co-rotating structure close to  the orbital plane,  \index{s}{Doppler tomography}Doppler tomography
of strong emission lines is the ideal tool at hand. Although providing
an image of the line emission distribution in  velocity space, and not
spatial   coordinates, \index{s}{spiral arms}spirals should be    readily identified as they
maintain their spiral shape in the velocity coordinate frame.

Since  the  application of   \index{s}{Doppler tomography}Doppler tomography to    the discs in CVs
(Marsh, this  volume), a range of  objects have been imaged, showing a
rich variety of accretion structures, but no clear evidence for \index{s}{spiral arms}spiral
arms in the  accretion discs. This was part  of the reason  that
the  interest in \index{s}{spiral arms}spiral waves   diminished after a  series of landmark
papers   in the  late  eighties   \cite{spruit1,spruit2}, and  a
solution   to    the angular  momentum  puzzle    was  sought in local
magneto-turbulent processes \cite{bh91}.   In 1997, however, convincing
observational evidence for tidally driven \index{s}{spiral arms}spirals  in the disc of a CV
was found for the first time.

\section{IP Pegasi in \index{s}{outburst}outburst}

As part of  a programme to study the  disc evolution of the \index{s}{cataclysmic variable!dwarf nova}dwarf nova
IP Pegasi\index{o}{IP Peg} through  \index{s}{Doppler tomography}Doppler tomography,  the  system was  observed just
after rise  to one  of its  \index{s}{outburst}outbursts.   Although such \index{s}{outburst}outbursts occur
regularly and have   a   characteristic recurrence time, one    cannot
predict them  accurately.   Obtaining scheduled telescope  time during
such  \index{s}{outburst}outbursts is therefore not  straightforward.  For IP Pegasi\index{o}{IP Peg}, the
average recurrence time for \index{s}{outburst}outbursts is 88  days, with a considerable
RMS variation of 18 days \cite{thesis}.

When a \index{s}{Doppler map}Doppler map was constructed of the hydrogen and helium emission
from IP Pegasi, a surprising emission pattern was found
\cite{shh97,shh98}.  The accretion disc was far from symmetric, instead \index{s}{accretion disc!emission}disc emission
was dominated by a two armed pattern in the lower left and upper right
quadrants of  the  \index{s}{Doppler map}Doppler map  (Figure  \ref{ip93}).  The \index{s}{spiral arms}spiral  arm
velocities range between 500 and 700 km/s,  corresponding to the outer
regions  of the accretion disc.   The  emissivity contrast between the
\index{s}{spiral arms}spirals and other parts of the disc, is about a  factor of $\sim$3 for
H$\alpha$,  and $\sim$5 in the case  of HeI6678  emission. There is no
evidence for line emission from the \index{s}{bright spot}bright spot. The \index{s}{spiral arms}spiral arm in
the lower quadrant, closest to the secondary, extends over an angle of
$\sim$100$^{\circ}$, and  is   weaker than the  arm in   the  opposite
quadrant. Strong  emission from the \index{s}{irradiation}irradiated  \index{s}{secondary star}secondary star is also
present, producing the prominent S-wave at low velocities.

\begin{figure}
\includegraphics[width=.4\textwidth]{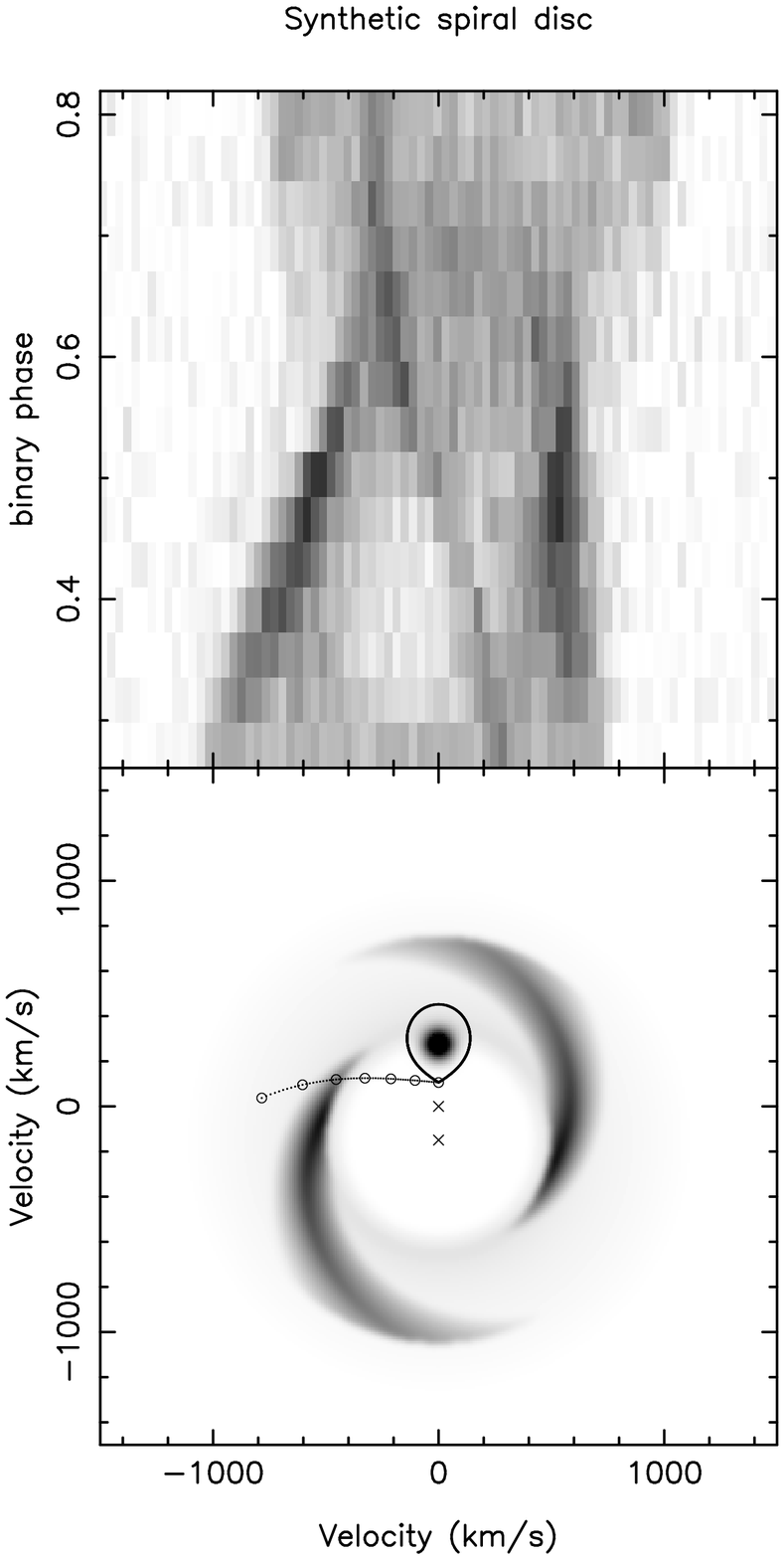}~\includegraphics[width=.4\textwidth]{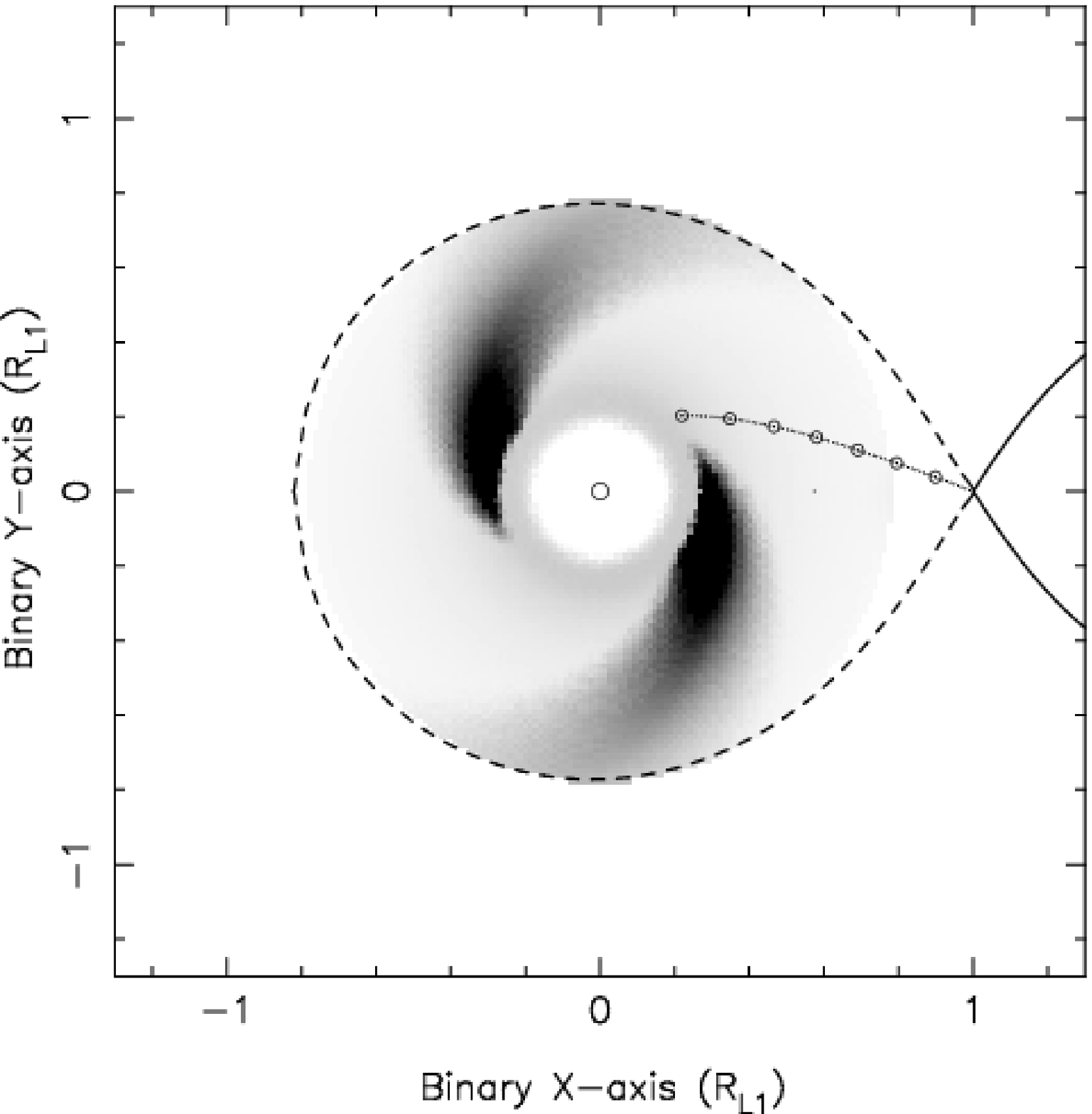} 
\caption[]{A simple model of a disc carrying a two armed \index{s}{spiral arms}spiral pattern. Bottom right the distribution of emission in XY coordinates, bottom left in Doppler coordinates and top is the predicted line profiles from such a disc. From \cite{shh97}}
\label{toymodel}
\end{figure}

The location of  this pattern  corresponds closely  to the radii   and
azimuths where tidally driven \index{s}{spiral arms}spiral waves  are expected.  Although the
Doppler  map\index{s}{Doppler map} is in  velocity coordinates,  and  not Cartesian position
coordinates,  a \index{s}{spiral arms}spiral  pattern in  position  space  corresponds  to a
\index{s}{spiral arms}spiral pattern in velocity space.   In Figure \ref{toymodel}, a simple
disc  model  is shown containing   an accretion disc  around the white
dwarf   with a purely    geometrical \index{s}{spiral arms}spiral  pattern  and a  \index{s}{Keplerian orbit}Keplerian
velocity field. The two armed trailing \index{s}{spiral arms}spiral in position coordinates,
maps into a similar two armed pattern in velocity space. Although this
model  is   a purely geometric  pattern, the   location of  the \index{s}{spiral arms}spiral
roughly corresponds  to the location  in the disc were  tidally driven
waves are expected  from model simulations.  Even  such a simple model
already  captures  most of   the  structure we   observe in IP Pegasi\index{o}{IP Peg},
indicating that most of the \index{s}{accretion disc!emission}disc emission in IP Peg during \index{s}{outburst}outburst is
indeed localised in a two armed pattern.

\begin{figure}
\centerline{\includegraphics[width=.5\textwidth,angle=-90]{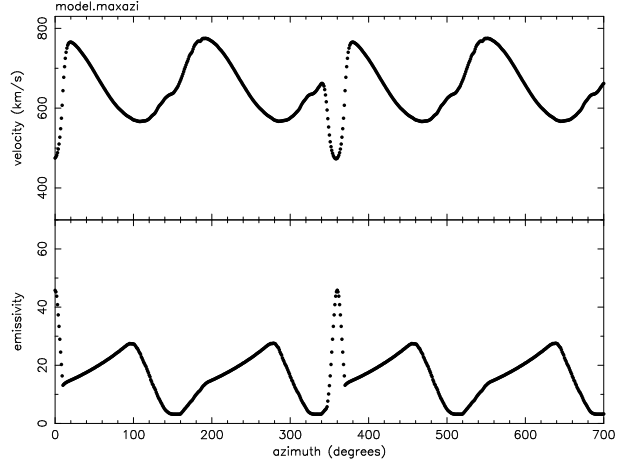} }
\caption[]{The azimuthal dependence of a two armed \index{s}{spiral arms}spiral produces a periodic variation in the velocity and emission strength of the \index{s}{spiral arms}spiral as a function of azimuth. The features near azimuth zero are due to the presence of companion star emission, which is also present in the observed data.}
\label{modelmaxazi}
\end{figure}

Let us look  at the main signatures  of \index{s}{spiral arms}spiral arms in the  observable
line profiles based on our simple model.  In Figure \ref{modelmaxazi},
the velocity and emissivity of the model disc is plotted as a function
of  azimuth in the Doppler  \index{s}{Doppler map}tomogram.  Azimuth  0 corresponding to the
positive V$_y$ axis, increasing clock wise.  The defining feature of a
trailing \index{s}{spiral arms}spiral is the change in velocity  as a function of azimuth in
the \index{s}{Doppler map}tomogram. This corresponds to a change in radius $R$ as a function of
azimuth   $\theta$ in position  coordinates that  is determined by the
opening  angle of the \index{s}{spiral arms}spiral, $dR/d\theta$.  The velocity field of the
disc relates this geometrical opening angle to  a particular velocity-azimuth relation  in the \index{s}{Doppler map}Doppler map.  In the \index{s}{spiral arms}spiral arm dominated line profiles, this  produces a regular motion   of the double   peaks, since
azimuth in position  coordinates translates  to  orbital phase in  the
observed spectrogram. Near orbital phases 0.25  and 0.75, the two arms
cross  over,  producing a sudden  jump  in the position  of the double
peaks.  For a symmetric \index{s}{Keplerian orbit}Keplerian disc on the other  hand, the line profile is symmetric and  the velocity  of the double peaks correspond  to the
velocity of the outer edge  of the disc.   This results in a  constant
double peak separation as  a function of orbital  phase and a circular
disc image in the  \index{s}{Doppler map}Doppler map.  With  a two armed \index{s}{spiral arms}spiral, the  double
peaks move,  varying their separation  considerably and sharply across
the  binary orbit in a  particular way.  Although an identification in
\index{s}{Doppler map}Doppler maps is perhaps more straightforward, spirals\index{s}{spiral arms} may thus also be
identified in the observed line profiles directly.

\begin{figure}
\centerline{\includegraphics*[width=.93\textwidth]{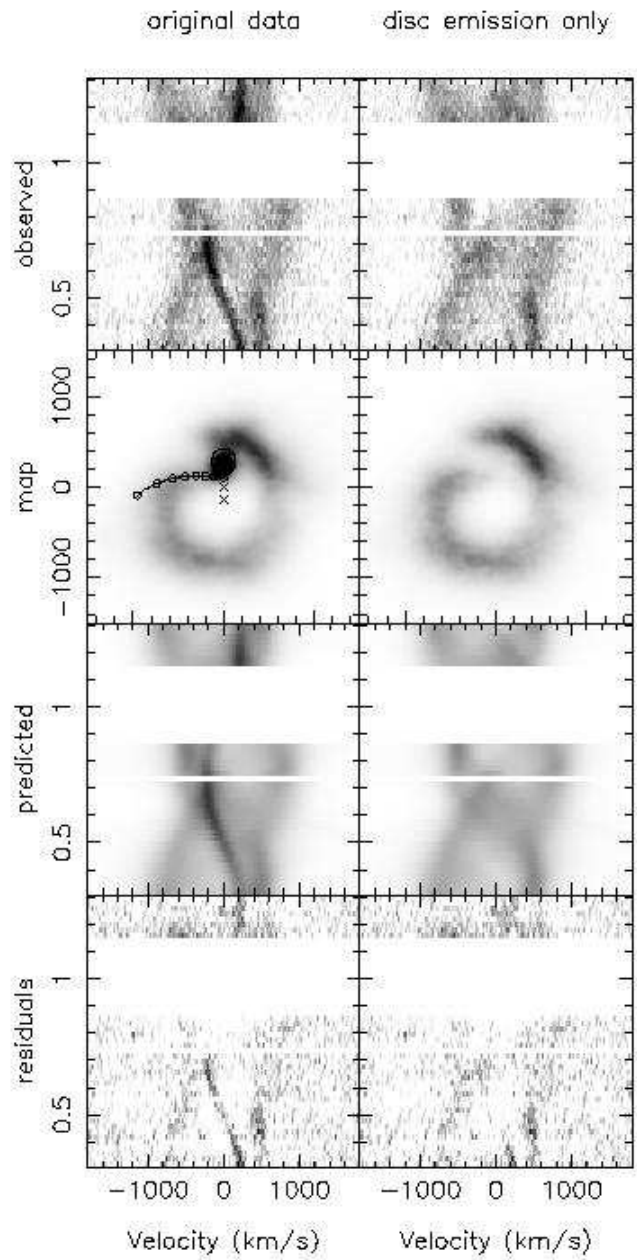} }
\caption[]{The \index{s}{spiral arms}spiral pattern persists throughout the \index{s}{outburst}outburst, above are HeI6678 data towards the end phases of \index{s}{outburst}outburst maximum. From top to bottom, observed data, \index{s}{Doppler map}Doppler map, predicted data and residuals. For the right hand panels, the contribution from the \index{s}{secondary star}secondary star has been subtracted. From \cite{thesis}}
\label{figipwht}
\end{figure}

A week later, during the same \index{s}{outburst}outburst, more spectroscopy was obtained
before  the decline of  the \index{s}{outburst}outburst  had started.   The Doppler image
(Figure  \ref{figipwht})  shows   that the   \index{s}{spiral arms}spiral pattern  persists
throughout the   \index{s}{outburst}outburst,   and the  secondary   star  now makes   a
considerably larger contribution (from 6\% to  10\%) to the line flux.
The  arm  in  the  upper  right quadrant  is still   stronger, and the
location  of the arms have not  changed, although the  upper right arm
appears shorter.  To  highlight the \index{s}{spiral arms}spiral arms  in the line profiles,
the S-wave of the \index{s}{secondary star}secondary star was subtracted from the data in the following manner. A Roche
lobe   shaped  mask was  used  to  select  all  the  emission from the
secondary,  which was then projected in  order to  produce the trailed
spectrogram of  the red star  emission.  This was subtracted  from the
observed data,  and  a Doppler image   of the disc   emission only was
constructed   (Figure   \ref{figipwht},  right).  The  dramatic  phase
dependence of the double peaks is now clearly visible, and is directly
related to the two \index{s}{spiral arms}spiral arms  in the disc.   This also confirms that
the presence of a strong S-wave due to the mass donor does not distort
the  reconstructed    disc  structure,  since  they   occupy different
locations in the velocity plane.
 
\begin{figure}
\centerline{\includegraphics*[width=.8\textwidth,angle=-90]{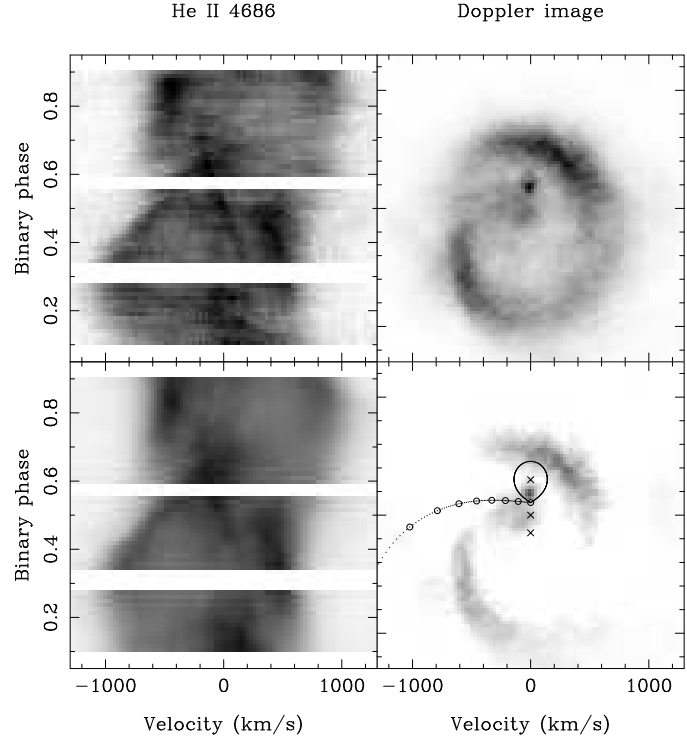} }
\caption[]{Ionised helium from the disc in IP Pegasi\index{o}{IP Peg} during the November 1996 \index{s}{outburst}outburst. The familiar \index{s}{spiral arms}spiral arms are present also in this line, together with emission from the front side of the secondary. Subtracting the symmetric component from the \index{s}{Doppler map}Doppler map, highlights the two armed spiral (bottom). Observed and predicted data are in the top left and bottom left panel respectively.}
\label{he4686map}
\end{figure}

\begin{figure}
\centerline{\includegraphics[width=.6\textwidth,angle=-90]{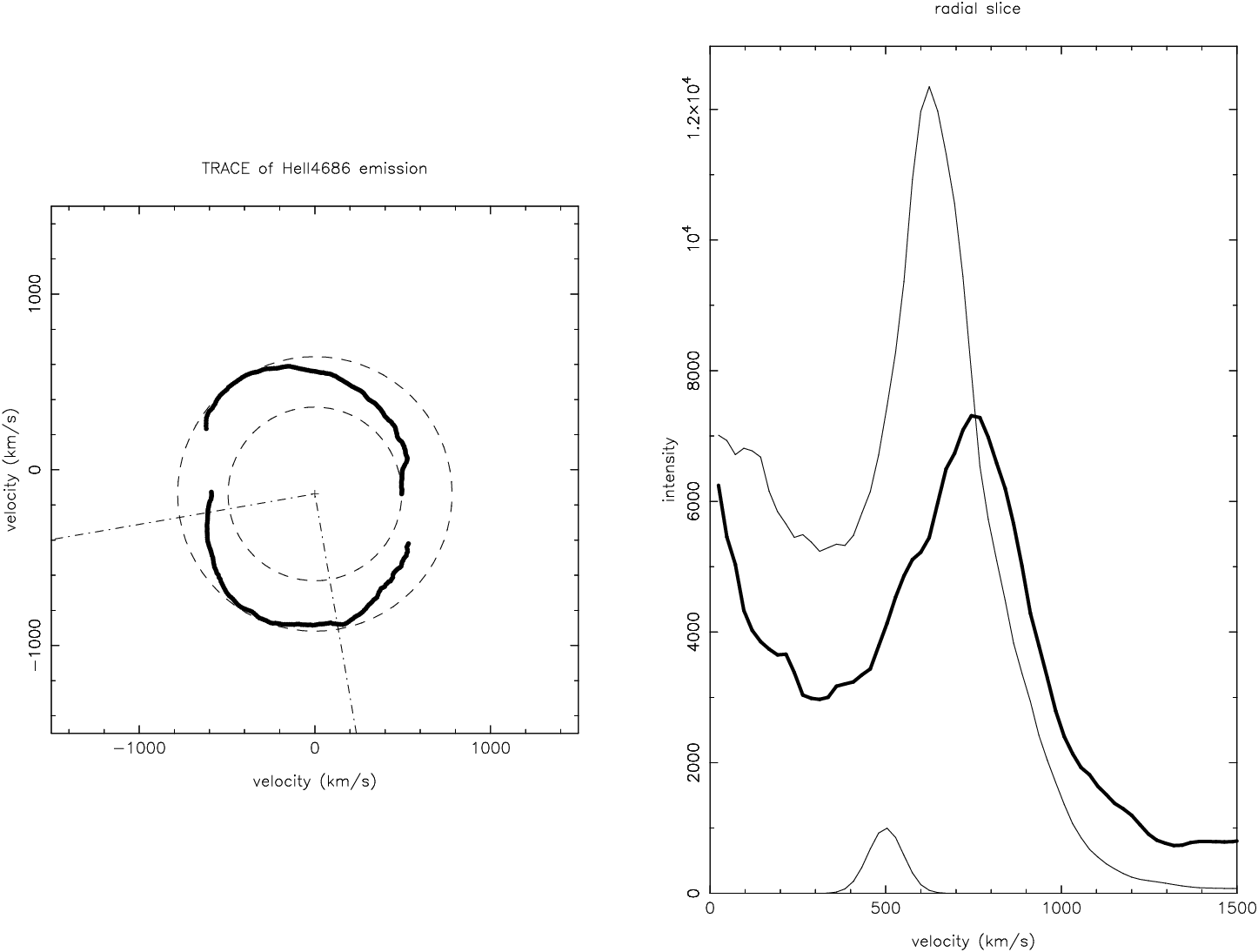} }
\caption[]{Tracing the \index{s}{spiral arms}spiral arms in the \index{s}{Doppler map}Doppler map through cross correlation with a Gaussian. Two radial slices through the \index{s}{Doppler map}Doppler map are plotted in the right panel, revealing the radial extent and position of the \index{s}{spiral arms}spiral arms. The instrumental profile (lowest curve) is plotted for comparison. The fitted positions of the \index{s}{spiral arms}spiral arms in the velocity plane are shown in the left panel. Circles denote the maximum/minimum velocity of the spirals and the two lines indicate the azimuth of the two slices plotted on the right.}
\label{maxazi}
\end{figure}

\subsection{Ionised Helium from the disc}

In a  different \index{s}{outburst}outburst Harlaftis et al. \cite{har99} secured a  whole orbit, including eclipse,
of   IP Pegasi\index{o}{IP Peg} with   high spectral  (40  km/s)  and  phase resolution
(0.01).  This time,  the \index{s}{outburst}outburst started  two days before
the observations, and the HeII emission line  at 4686\AA~ was observed
to provide a comparison with the previously observed H$\alpha$ and HeI
emission patterns.  The trailed spectrogram of HeII (Figure
\ref{he4686map}),   again shows the  familiar  behaviour of the double
peaks from the  disc, leading to  a  two armed  \index{s}{spiral arms}spiral in  the Doppler
\index{s}{Doppler map}tomogram. A  very   similar emission pattern from  the   disc was also
present in the nearby Bowen blend consisting of CIII/NIII emission and
the Helium  I line at 4471\AA. The  symmetric component of the map was
calculated, using  the velocity of  the  \index{s}{white dwarf}white dwarf  as the centre of
symmetry, and subtracted in order to highlight the location of the two
arms (Figure \ref{he4686map}, bottom right).

In  order to characterise the  properties of the \index{s}{spiral arms}spirals quantitatively,
we determined  the position of  the two \index{s}{spiral arms}spiral arms  as a  function of
velocity in  the reconstructed  \index{s}{Doppler map}tomogram.  A  slice  through  the map,
starting from  the \index{s}{white dwarf}white dwarf at $V_y=-137$  km/s, was  made for each
azimuth in  order to extract  the radial profile.  Each radial profile
was then  cross correlated with a Gaussian  in order  to determine the
velocity of the spiral\index{s}{spiral arms} at that particular azimuth. Figure \ref{maxazi}
plots  the  fitted positions  of the  \index{s}{spiral arms}spirals   in the velocity  plane
together with two radial slices across the map. The two \index{s}{spiral arms}spirals can be
traced for  almost 180 degrees, and  the  velocity of the  arms varies
from 495 km/s to 780 km/s, indicated by  the two dashed circles.  Near
the  azimuths where the \index{s}{spiral arms}spiral   cross  over, the fitting  assumptions
break down as there is a sudden transition between the velocity of one
arm and the next. Unfortunately, the contrast of the  arms is too low
to   follow this switching  with  our  cross  correlation method, or a
double  Gaussian.  The  widths of  the  arms are significant, and  even
change as a function of azimuth. We  are thus resolving an intrinsically
broad  feature. Such an  analysis  will  clearly profit from
even higher resolution and signal to noise data since at peak intensity the
width  of the  arm  is comparable to  our resolution  element, and the
maximum \index{s}{entropy}entropy constraint will tend to broaden features as much as is
allowed by the signal to noise of  the data.

A tidally distorted disc
will not have a pure \index{s}{Keplerian orbit}Keplerian  velocity field, and  we will see later
that the  observations  indeed indicate  this is  not  the case  in IP
Pegasi\index{o}{IP Peg}.  However,  we can estimate  the   radii corresponding  to  the
velocity of the \index{s}{spiral arms}spirals under the assumption of \index{s}{Keplerian orbit}Keplerian velocities. The arms
then cover a substantial part  of the disc, between  0.3 and 0.9 times
the \index{s}{distance}distance to the $L_1$ point,  with the strongest emission from the
outer regions. The dynamics of the majority of the accretion disc material are thus affected by the presence of these \index{s}{spiral arms}spirals.

\begin{figure}
\centerline{\includegraphics[width=.6\textwidth,angle=-90]{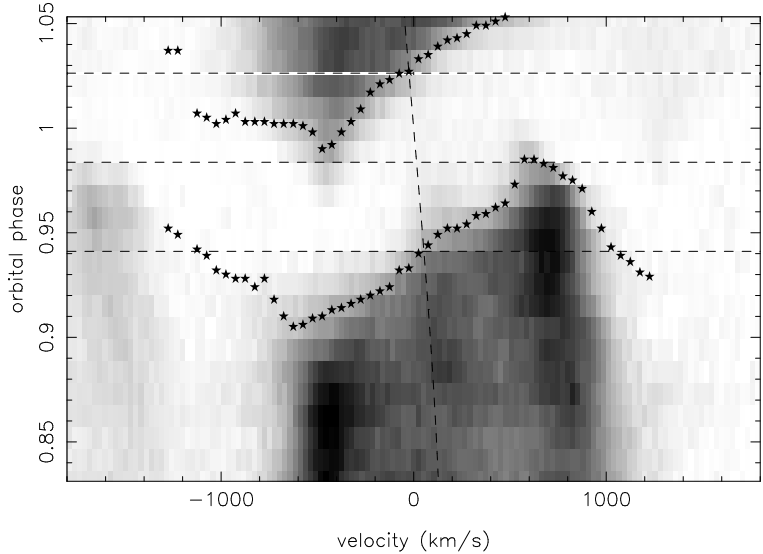} }
\caption[]{Measurements of the phases of half depth across the HeII eclipse are denoted by asterisks. Horizontal, dashed lines are the average post-eclipse half depth (top), pre-eclipse half-depth (bottom) and mid-eclipse. The orbital phases are based on the conjunction of the \index{s}{white dwarf}white dwarf.}
\label{zwave}
\end{figure}

The  eclipses of the  lines are  also  affected by the large accretion
disc asymmetry.  In Figure \ref{zwave},  the eclipse of  the Helium II
emission line  is plotted. For each  velocity bin, the  orbital phases
where half the light is eclipsed are  marked by asterisks.  The outer disc
on the blue-shifted peak is   eclipsed first, followed by emission  at
higher, blue-shifted velocities. After the  blue side  of the disc  is
covered, the  red-shifted  is emission is  progressively eclipsed, and
during \index{s}{egress}egress, the situation  reverses. This is the  classical pattern
of the eclipse of  a pro-gradely rotating disc\index{s}{accretion disc!emission},  where the disc velocity
increases with decreasing  radius. A strictly \index{s}{Keplerian orbit}Keplerian accretion disc
would result in a  smooth eclipse, symmetric  around orbital phase  0.
Mid-eclipse for  the  \index{s}{accretion disc!emission}disc emission  from IP  Peg\index{o}{IP Peg} occurs  considerably
earlier  (orbital phase 0.987) compared to  the continuum, and most of
the  light is  eclipsed  well  before  \index{s}{white dwarf}white dwarf  \index{s}{ingress}ingress.   A large
asymmetry in the  outer disc is  thus corroborated by the eclipse. The
distorted disc  eclipse is a   combination of the spiral asymmetry  as
well as deviations from \index{s}{Keplerian orbit}Keplerian velocities.  Although these
eclipse phases  are not fitted by  \index{s}{Doppler tomography}Doppler tomography codes, high  quality eclipse data  has the
potential   to  reveal  departures   from  \index{s}{Keplerian orbit}Keplerian  velocities   and
\index{s}{disentanglement}disentangle these two effects.

Eclipse mapping of this data in  both the lines and continuum, reveals
the presence  of a two  armed  asymmetry in  the outer  regions of the
disc, at the azimuths corresponding to the \index{s}{spiral arms}spiral  arms in the Doppler
maps \cite{bhs00} (see also Baptista, this volume).  The velocities of
the  \index{s}{spiral arms}spirals as derived from the  emission lines, can be compared with
the position of the \index{s}{spiral arms}spirals from the eclipse  map. This indicates that
velocities  near the   \index{s}{spiral arms}spirals  deviate considerably  from   the local
\index{s}{Keplerian orbit}Keplerian value. The two imaging  methods thus nicely complement  each
other and both support a tidal interpretation of the spirals\index{s}{spiral arms}.

Apart  from    the  data  sets    discussed  here,   Morales-Rueda  et
al. \cite{luisa}, also recovered prominent \index{s}{spiral arms}spiral waves in the disc of
IP Pegasi\index{o}{IP Peg} during \index{s}{outburst}outburst in a range of emission  lines. The
disc structure during their observations, about 5 days after the start
of the  \index{s}{outburst}outburst, is very similar to  the \index{s}{Doppler map}Doppler maps presented here,
with a  slightly stronger arm  in the upper  right quadrant. They also
note   the  shifted  emission  line   eclipses and  shielding  of  the
\index{s}{irradiation}irradiated companion star by the geometrically thick disc\index{s}{accretion disc!thickness}.

\index{s}{Doppler tomography}Doppler tomography of IP Pegasi\index{o}{IP Peg}  during \index{s}{outburst}outburst thus invariably shows
the presence of \index{s}{spiral arms}spiral shaped  disc asymmetries.  The two armed spiral\index{s}{spiral arms}
dominates the \index{s}{accretion disc!emission}disc emission from the start of the \index{s}{outburst}outburst maximum and
persists for at least 8 days, corresponding to about 50 binary orbits.
The \index{s}{spiral arms}spirals  are present  in a  range of  emission lines  from neutral
hydrogen   to \index{s}{ionisation}ionised helium,  the  latter   indicating  that the  gas
concerned has to be hot, although it is not clear if we are looking at
direct emission from  the  shock, or  recombination  emission from the
\index{s}{spiral arms}spiral arms.  The asymmetry between the two  arms that was observed in
the  discovery data,  is also present   at  other epochs and in  other
lines.  The  disc structure   is    co-rotating with  the binary,    and
corresponds to   the  velocities were  tidally  driven  \index{s}{spiral arms}spiral  arms are
expected. A more detailed  comparison between theory and  observations
is discussed  by Boffin, this volume. We  shall  see that the observed
properties of the \index{s}{spiral arms}spirals   fit recent hydro-dynamical  simulations of
such discs in detail \cite{ss99}.

\begin{figure}
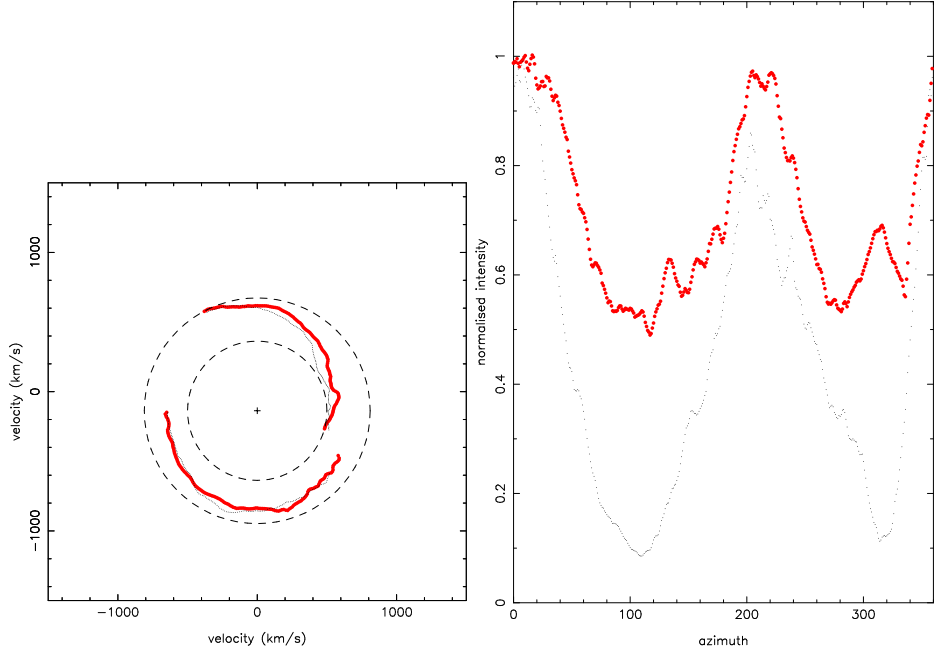

\centerline{\includegraphics[width=.5\textwidth]{figure8a.ps}~\includegraphics[width=.5\textwidth]{figure8b.ps} }
\caption[]{Comparing the properties of the \index{s}{spiral arms}spirals in two emission lines from the same data set. The position and intensity of the spirals were traced as a function of azimuth in the H$\alpha$ (thick line) and HeI6678\AA~ \index{s}{Doppler map}tomograms constructed from NTT echelle spectroscopy during the August 1999 \index{s}{outburst}outburst of IP Peg\index{o}{IP Peg}. The intensities are normalised with respect to the maximum emissivity values of each line. }
\label{NTT}
\end{figure}

Although  the \index{s}{accretion disc!emission}disc emission  of the different  data sets is in general
terms the same, differences between the  various data sets should tell
us  something about the  evolution of the  \index{s}{spiral arms}spiral across the \index{s}{outburst}outburst (and therefore their origin). In addition, comparing  lines with different excitation  potential allows us to
investigate  the     physical  conditions   of     the  emitting   gas
directly. Unfortunately, the various data sets have been obtained with
different  telescopes   and  instruments,  different  resolutions, and
different  orbital  phase  coverage.   Disentangling those  systematic
effects  from  true  variations of  the  \index{s}{spiral arms}spiral  in  various  lines is
therefore not  straightforward.  Ideally one would   observe a range of
emission lines simultaneously with the  same instrumental setup, while
covering  a  substantial part  of   the disc  \index{s}{outburst}outburst  to  study  its
evolution.  As a first step,  we have obtained echelle spectroscopy of
IP Pegasi\index{o}{IP Peg} during a recent \index{s}{outburst}outburst with the NTT (Steeghs \& Boffin, in
preparation).  Covering a large part of  the optical  spectrum at high
resolution, ensures that we will be able  to compare the properties of
the  \index{s}{spiral arms}spiral  among a   large set of  emission   lines,  obtained under
identical circumstances.  Figure  \ref{NTT} compares the properties of
the \index{s}{spiral arms}spirals between just two lines covered, and  indicates a  very similar
position  for the \index{s}{spiral arms}spirals in  terms  of velocity  as   a function of   azimuth for the
H$\alpha$ and HeI6678\AA~    lines. The intensity modulation,   on the
other hand is significantly different.

\section{The quiescent disc}

In  stark  contrast with  the  prominent \index{s}{spiral arms}spiral waves  observed during
\index{s}{outburst}outburst,  \index{s}{Doppler map}Doppler mapping of  IP Pegasi  during quiescent epochs does
not  reveal   such features   \cite{mh90,ksh94,hmd94,wolf98}. Although
significant disc asymmetries   are observed even during \index{s}{quiescence}quiescence,   it is
clear that the open \index{s}{spiral arms}spiral pattern the disc carries in \index{s}{outburst}outburst is not
present.  The disc radii  of \index{s}{cataclysmic variable!dwarf nova}dwarf novae are expected to
vary significantly across the \index{s}{outburst}outburst  cycle. In the disc instability
picture, a \index{s}{cataclysmic variable!dwarf nova}dwarf nova disc is in a  cool state during quiescent phases, when most mass transferred from the mass donor  is not accreted by the
white  dwarf.   The disc  density  builds up until   a radical opacity
change  due   to the  \index{s}{ionisation}ionisation of   hydrogen flips  the  disc from a
neutral, cool state to a hot, \index{s}{ionisation}ionized state. Angular momentum transport
in this hot state is increased by a factor of 10  or so, which results
in a rapid expansion of the disc at the onset  of the \index{s}{outburst}outburst. During
\index{s}{outburst}outburst,  the disc   is depleted as  more  mass  is accreted than  is
transferred from the  mass donor, until the density in the disc drops below the critical value again and   the  system returns to \index{s}{quiescence}quiescence.

Disc  radius\index{s}{accretion disc!radius} variations are  indeed observed in   \index{s}{cataclysmic variable!dwarf nova}dwarf novae, and the
disc in IP Pegasi\index{o}{IP Peg} fills between half  of the primary \index{s}{Roche equipotential}Roche lobe during
quiescence   up    to   most      of    the  Roche     lobe      during
\index{s}{outburst}outburst \cite{wood86,wmh93,thesis}. Tidally driven waves are thus much
more likely to have an effect on the disc in \index{s}{outburst}outburst, since the tidal
torques are a steep function  of the \index{s}{distance}distance to the \index{s}{secondary star}secondary star.  The fact
that \index{s}{spiral arms}spirals  are  not prominent  in \index{s}{quiescence}quiescence,  is therefore what one
expects if they  are due to the  tidal torques of the  companion star\index{s}{secondary star} \cite{ss99}.
The detailed properties  of the \index{s}{spiral arms}spirals   depend upon disc  temperature,
radius  and the  \index{s}{mass ratio}mass ratio  of the   binary.   The most  likely discs
carrying prominent open-armed \index{s}{spiral arms}spirals are thus \index{s}{cataclysmic variable!dwarf nova}dwarf novae in \index{s}{outburst}outburst
and the  \index{s}{cataclysmic variable!nova-like}nova-like variables,  where the disc  is  always in  the hot,
\index{s}{ionisation}ionised state.  Dwarf   novae  are likely  to display  highly  variable
\index{s}{spiral arms}spiral arm  structures, depending on the  state of the system, whereas
\index{s}{cataclysmic variable!nova-like}nova-like variables would   provide  a more persistent  tidal  pattern
since large \index{s}{accretion disc!radius}disc radius variations are relatively rare.

Marsh, this volume,   has compiled an up to date   overview of all   the
systems  for which \index{s}{Doppler map}Doppler maps  have been published. One would perhaps
expect that  such \index{s}{spiral arms}spirals should  have been found  before if they are a common phenomenon. The number   of Doppler  maps  of dwarf   novae in \index{s}{outburst}outburst   are very rare
indeed. And  even in \index{s}{quiescence}quiescence, some  well known systems  have not yet
been \index{s}{Doppler map}mapped using good quality data.  Unfortunately, in the
case  of \index{s}{cataclysmic variable!nova-like}nova-likes where  one would have  the potential of observing a
persistent \index{s}{spiral arms}spiral  pattern, \index{s}{Doppler map}Doppler mapping   has shown complicated
emission geometries. In those systems, line emission  is  not  dominated by a disc component, instead prominent spots dominate the \index{s}{Doppler map}tomograms whose interpretation   is  unclear. However, IP
Pegasi\index{o}{IP Peg} is no longer the only system with evidence for distortions of a
spiral nature. The next section will  discuss a small number of other
systems which display very similar disc behaviour.

\section{Disc asymmetries in other systems}

\subsection{EX Dra}
\begin{figure}
\centerline{\includegraphics*[width=.6\textwidth,angle=+90]{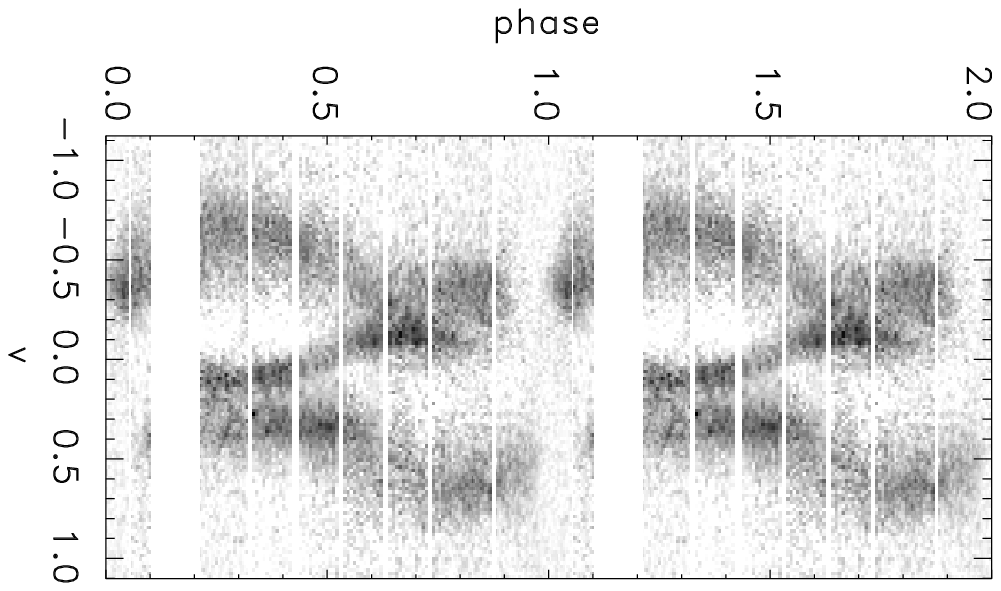}~\includegraphics*[width=.6\textwidth]{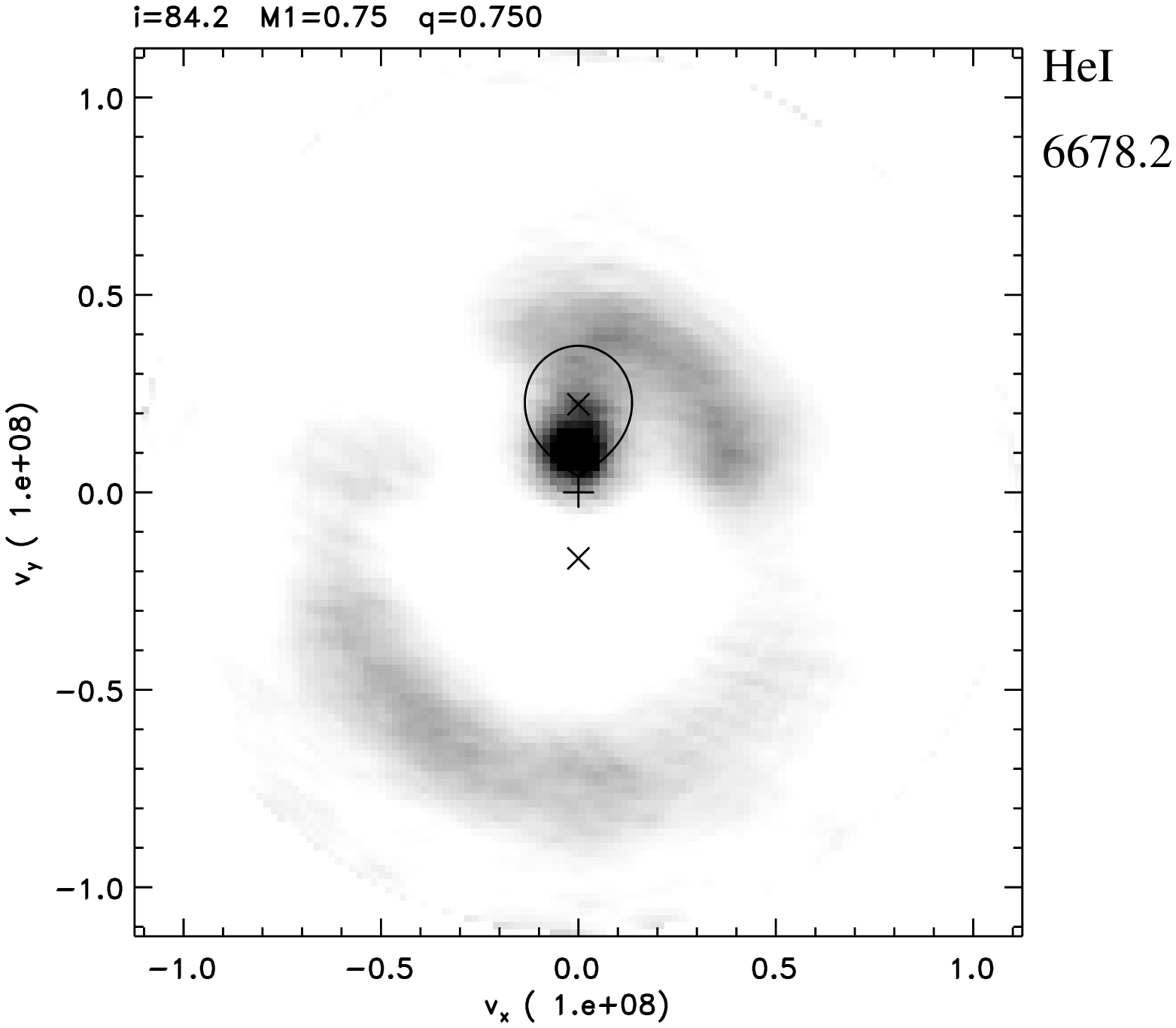} }
\caption[]{\index{s}{Doppler tomography}Doppler tomography of EX dra\index{o}{EX Dra} in \index{s}{outburst}outburst \cite{exdraspiral}. Left the observed HeI 6678\AA~ emission and its \index{s}{Doppler map}tomogram on the right. Apart from the \index{s}{irradiation}irradiated secondary, the disc contains a two armed asymmetry with a shorter arm in the top right and a more extended arm in the lower left quadrant.}
\label{exdramap}
\end{figure}

The dwarf  nova  EX Dra\index{o}{EX Dra} was the  second  object to show  similar  disc
behaviour.  This is another eclipsing system above  the \index{s}{period gap}period gap and
is in many ways very similar to  IP Pegasi \cite{bill,fiedler}.  Joergens
et  al. \cite{exdraspiral}  present  Doppler images  of \index{o}{EX Dra}EX  Dra during
\index{s}{outburst}outburst,  and the  similarity of the  \index{s}{Doppler map}Doppler maps  with  those of IP
Pegasi  is obvious (Figure  \ref{exdramap}).  Although perhaps not  as
convincing as in IP Pegasi\index{o}{IP Peg} in all the mapped lines, the disc carries a
two armed asymmetry   during \index{s}{outburst}outburst  in the  same  quadrants of  the
\index{s}{Doppler map}Doppler map.  The appealing characteristic of this  object is its very
short  \index{s}{outburst}outburst recurrence time. Although  its orbital period and mass
ratio is  very similar to IP  Peg, EX Dra's  \index{s}{outburst}outbursts recur  every 23
days, and catching such a system in \index{s}{outburst}outburst  is thus more likely.  It
would be an ideal  target for a longer  campaign, with the prospect of
securing the disc   evolution across  the  whole \index{s}{outburst}outburst   cycle, and
performing time lapsed tomography in order to construct a movie of the
accretion   disc  along its  \index{s}{outburst}outburst  cycle.  A  photometric campaign
lasting  over 6 weeks was conducted   during the summer  of 2000, just
after this workshop. Although  no \index{s}{Doppler tomography}Doppler tomography will be possible,
two \index{s}{outburst}outburst  cycles were  covered,  and \index{s}{eclipse map}\index{s}{eclipse mapping}eclipse mapping methods  will
reveal the  development   of disc asymmetries   and radial temperature
variations.

\subsection{U Gem}

\begin{figure}
\centerline{\includegraphics[width=.6\textwidth]{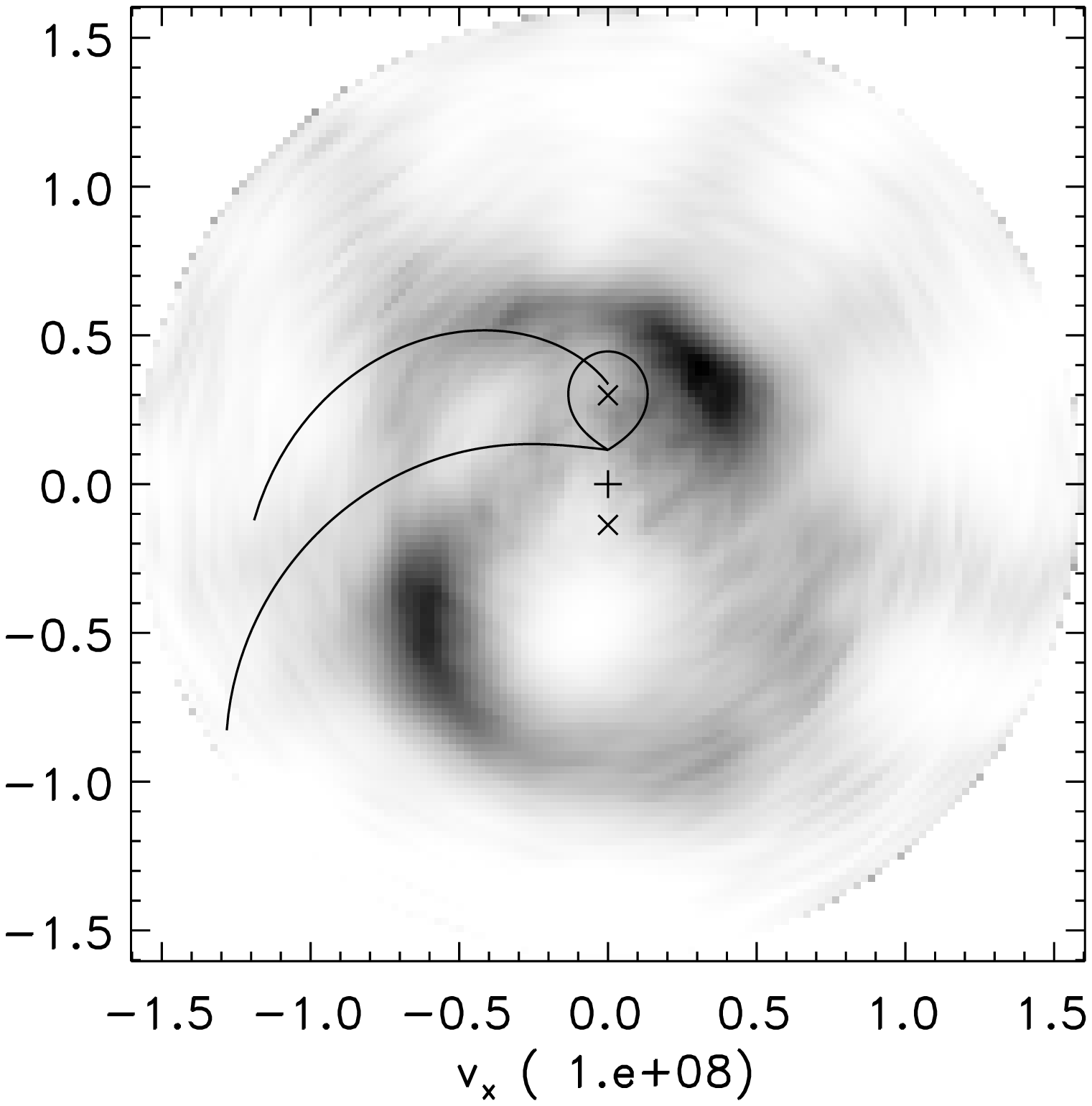}}
\vspace{.5cm}
\caption[]{Doppler \index{s}{Doppler map}tomogram of the HeII 4686\AA~ emission from U Gem\index{o}{U Gem} in \index{s}{outburst}outburst \cite{groot}. Two striking \index{s}{spiral arms}spiral arms are revealed, matching the properties of a tidal spiral pattern. }
\label{ugem}
\end{figure}

U  Gem\index{o}{U Gem}  is one  of  the brightest and   best studied  CVs,  reaching a
magnitude  of V$\sim$8.5 during  \index{s}{outburst}outburst.  Surprisingly,  no \index{s}{outburst}outburst
Doppler  tomography of this system was  available until very recently,
Groot   observed  the  system    during    the March   2000   \index{s}{outburst}outburst
\cite{groot}. Figure \ref{ugem} shows  the HeII \index{s}{Doppler map}tomogram  of U  Gem in
\index{s}{outburst}outburst, revealing a prominent  two armed  \index{s}{spiral arms}spiral pattern.
  The \index{s}{spiral arms}spirals are  strong even towards  the late phases of  the
\index{s}{outburst}outburst maximum, and weaken once decline sets in. This supports
a  tidal  interpretation for these  structures  since  the \index{s}{spiral arms}spirals are
there when the disc is hot and large, but weaken once the disc shrinks
back towards  \index{s}{quiescence}quiescence like in IP Pegasi. The rise and  decline phases of the \index{s}{outburst}outbursts
are particularly useful for testing the tidal interpretation of \index{s}{spiral arms}spiral
waves.  During  those  phases  the  \index{s}{spiral arms}spiral  arm  geometry will  change
dramatically, and the  exact nature of the  evolution will tell us  if
indeed the \index{s}{spiral arms}spirals behave as tidally driven \index{s}{spiral arms}spirals  would do, and how
they affect the \index{s}{angular momentum}angular momentum exchange in the accretion disc.

\subsection{SS Cyg}

\begin{figure}
\centerline{\includegraphics*[width=.65\textwidth,angle=-90]{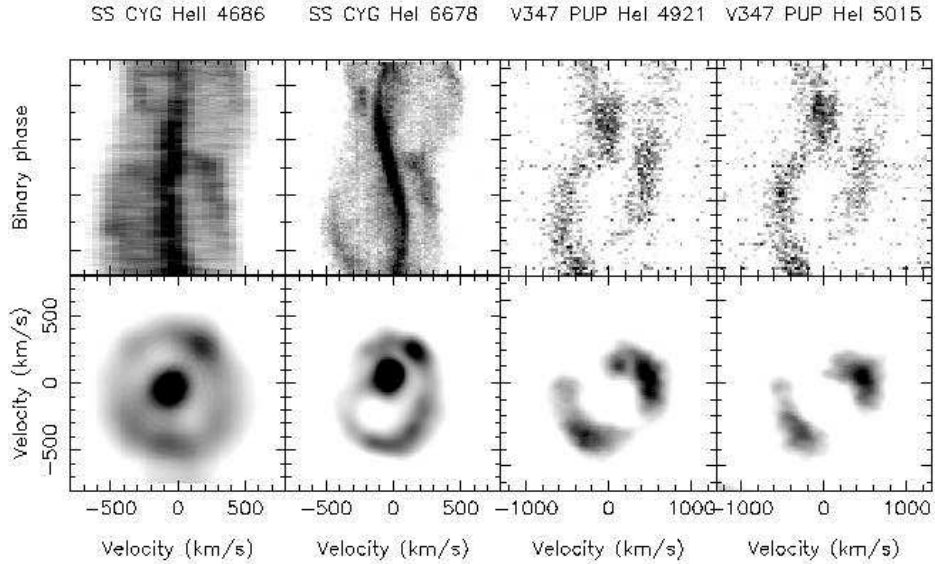}}
\vspace{.5cm}
\caption[]{Two other cases of distorted accretion discs with a two armed asymmetry. Left the HeII and HeI emission of SS Cygni\index{o}{SS Cyg} during \index{s}{outburst}outburst, right two panels the eclipsing \index{s}{cataclysmic variable!nova-like}nova-like V347 Pup\index{o}{V347 Pup}.}
\label{others}
\end{figure}

During \index{s}{outburst}outburst, SS  Cygni\index{o}{SS Cyg} is the   brightest CV in  the sky,  with an
almost continuous  light curve available for the  last 100 years thanks
to amateur astronomers. Its \index{s}{outburst}outburst history has  been well studied and
compared   with  disc  instability   predictions \cite{canizzo}.  Phase
resolved spectroscopy reveals emission from the \index{s}{secondary star}secondary star as well
as a distorted accretion disc, both in \index{s}{quiescence}quiescence and \index{s}{outburst}outburst
\cite{pais1,pais2}.     \index{s}{Doppler tomography}Doppler tomography  during   \index{s}{outburst}outburst revealed
enhanced emission from   the disc in  the   top right and   lower left
quadrant  of  the   \index{s}{Doppler map}tomograms  in  the  Balmer, HeI   and   HeII lines
\cite{slingpaper}.  \index{s}{maximum entropy}MEM maps  of two emission  lines are presented  in
Figure \ref{others}.   Again, the disc  has a two armed asymmetry, but
on the other hand differs significantly from a simple two armed \index{s}{spiral arms}spiral
pattern.  SS Cyg\index{o}{SS Cyg} is  an intermediate \index{s}{inclination}inclination  system, and our line
of  sight thus penetrates  deeper into  the disc atmosphere, producing
broad   absorption  wings  from   the   optically   thick disc  during
\index{s}{outburst}outburst. This may complicate the comparison  between models of \index{s}{spiral arms}spiral
arms and the line emission from them  because of more intricate, and
poorly   understood,  radiative  transfer conditions in the vertically stratified disc.   However,  these
differences need  an explanation and given  the brightness  of SS Cyg,
justifies high resolution  phase resolved spectroscopy  of this system
in \index{s}{quiescence}quiescence and \index{s}{outburst}outburst.

\subsection{V347 Pup}

The accretion  discs of \index{s}{cataclysmic variable!nova-like}nova-likes are  in many ways regarded as dwarf
nova in  permanent \index{s}{outburst}outburst, since the high  mass transfer rate in the
disc keeps  the  disc  gas  \index{s}{ionisation}ionized and stable  against   thermal disc
instabilities. Unfortunately the properties  of the emission lines  of
these systems remains a puzzle. The atlas of \index{s}{Doppler map}Doppler maps from
\cite{ksh94} contains many \index{s}{cataclysmic variable!nova-like}nova-like  systems, for which evidence
for  an  accretion disc is  absent in  most cases.    One of the few
\index{s}{cataclysmic variable!nova-like}nova-likes that  appears to reveal a clear  signature  of an accretion
disc is the eclipsing \index{s}{cataclysmic variable!nova-like}nova-like V347 Pup\index{o}{V347 Pup} \cite{still}. \index{s}{Doppler map}Doppler mapping
revealed the  presence  of an accretion disc,   containing a two armed
asymmetry.  However,  the properties of some of the  emission lines at a different 
epoch several years later were quite different (Steeghs   et  al., in preparation).    The
\index{s}{Balmer line}Balmer lines were dominated  by a spot of emission  in the  lower left
side   of the \index{s}{Doppler map}tomogram  rather  than  the disc,  a typical  feature of
\index{s}{cataclysmic variable!nova-like}nova-like variables.  However,  the HeI  emission (Figure \ref{others}) appears
to be  originating   from the disc   only, and  displays a  two  armed
asymmetry like that  was observed in the  \index{s}{Balmer line}Balmer lines by Still et al. \cite{still}
The opportunities  to investigate the tidal  structures of a high mass
transfer rate  accretion disc are ideal in  this eclipsing object, and
follow up NTT data is under investigation.

\subsection{A systematic picture?}

Phase resolved spectroscopy of CVs  in \index{s}{outburst}outburst, suitable for  Doppler
mapping of the accretion disc,  is still limited to  a small number of
objects.  It is very encouraging that on  the occasions that such data
are available  and  can   be compared,  very  similar  accretion  disc
distortions are seen.  This suggests a common physical cause for these
large scale  asymmetries and tidally driven waves appear to
offer the  most likely explanation.   It is interesting  to note that all
the \index{s}{cataclysmic variable!dwarf nova}dwarf novae  that have indicated  spiral-type distortions are long
period systems above the \index{s}{period gap}period gap.  One does expect the structure of
the \index{s}{spiral arms}spiral arms to depend on the \index{s}{mass ratio}mass ratio of the binary, with higher
mass   ratio binaries corresponding to  a  heavier \index{s}{secondary star}secondary star that
induces stronger  tidal torques.  On  the other hand selection effects
work  against us since  the \index{s}{outburst}outburst  of the short  period systems are
usually  short, and suitable  \index{s}{outburst}outburst  spectroscopy  of short  period
\index{s}{cataclysmic variable!dwarf nova}dwarf novae during \index{s}{outburst}outburst maximum is  extremely rare.  Tomography of
OY Car\index{o}{OY Car} in  \index{s}{outburst}outburst \cite{oycar}, reveals an extended  arm on the side
of the \index{s}{gas stream}gas stream impact, but not  a two armed \index{s}{spiral arms}spiral pattern. Limited
resolution and signal to noise in this case prevents a strong case for
or against \index{s}{spiral arms}spirals in SU UMa systems.

The important question is whether \index{s}{spiral arms}spiral waves affect the structure of
accretion discs under a   wide range of   conditions, or are merely  a
dynamical   side effect  of  the   expanding disc  during  \index{s}{cataclysmic variable!dwarf nova}dwarf novae
\index{s}{outburst}outbursts.  Clear   progress requires  a more   balanced observational
picture covering the observational parameter space in terms of orbital
periods, \index{s}{mass ratio}mass ratios,  \index{s}{outburst}outburst behaviour, etc.    This relies on  our
ability to obtain a considerable   number of data sets of  \index{s}{outburst}outbursting
dwarf    novae,   in particular     during    the  rise and    decline
phases. Flexible, or even robotic,  scheduling of telescope time would
be highly beneficial to such projects.

\index{s}{Doppler tomography}Doppler tomography  has  proven  to be   an invaluable tool   for  the
discovery and study  of \index{s}{spiral arms}spiral waves in the  discs of CVs. With better
resolution  and signal to noise   data  becoming available with  large
aperture  telescopes, however, there  is also room for improvements to
the technique itself.   In the next section  I describe an extension to
the  \index{s}{Doppler tomography}Doppler tomography technique that aims  to improve our ability to
fit  to data sets  containing anisotropic  emission sources, a  rather
common situation.

\section{Modulation mapping}

\index{s}{Doppler tomography}Doppler tomography provides a velocity resolved image of the accretion
flow in the corotating frame of the binary. Since it relies on only a
few  basic assumptions,  such   images can  be  recovered  in a  model
independent way. They then provide a  perfect frame in which to compare
data with models.
One of those assumptions is that the flux  from any point fixed in the
rotating frame is  constant (Marsh, this volume). However, observations
show the presence of  anisotropic  emission sources, that modulate  their
emission as a  function of the orbital phase.   Some general examples are
the  \index{s}{irradiation}irradiated  front of  the mass   donor,   the \index{s}{bright spot}hot spot,   and the
\index{s}{anisotropic radiation}anisotropic emission from \index{s}{spiral arms}spiral  shocks. \index{s}{Doppler tomography}Doppler tomography can  still
be used in that case, since the  \index{s}{Doppler map}Doppler map serves  to present a time
averaged image of the distribution of  line emission. However, one will
not be  able to  fit   the data very  well,  and  the phase  dependent
information contained  in the  observed   line profiles  is lost.  The
remainder of  this review will present  a straightforward extension to
the \index{s}{Doppler tomography}Doppler tomography method that tries to remove the above mentioned
assumption from tomography.

\subsection{Extending \index{s}{Doppler tomography}Doppler tomography}

\begin{figure}
\centerline{\includegraphics[width=1.\textwidth]{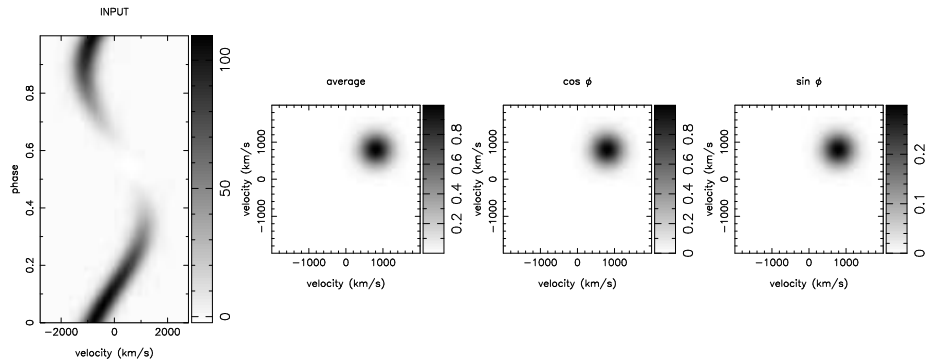}}
\caption[]{A spot that modulates its emission across the orbital phase is not only characterised by its position in the $V_xV_y$ plane but also the amplitude and phase of its modulation. Two additional velocity images are used to store the cosine and sine amplitudes of such S-waves. All modulations on the orbital periods can then be described as the sum of the image values in the three velocity images.}
\label{spot}
\end{figure}

Rather than assuming that the flux from each point in the binary frame
is constant,  I include  modulations   of the   flux on  the   orbital
period. The line flux from each location (in terms  of its position in
the $V_xV_y$ plane of the  \index{s}{Doppler map}tomogram) is not  just characterised by the
average line flux, but also the amplitude  and phase of any modulation
on the orbital period (Figure \ref{spot}). We thus have two additional
parameters  to  describe the  line   flux  form a   specific  velocity
vector. The flux $f$ from a modulated S-wave as a function of the orbital phase $\phi$ can be written as;
\[ f(\phi) = I_{avg} + I_{cos}\cos{\phi} + I_{sin}\sin{\phi}
\]
\noindent 
where $I_{avg}$ is the average line flux for the \index{s}{s-wave}s-wave, and $I_{cos}$
and $I_{sin}$ the cosine  and sine amplitudes (Figure \ref{spot}).  In
other   words,    the      amplitude  of        the    modulation   is
$\sqrt{I_{cos}^2+I_{sin}^2}$,        and         its       phase    is
$\tan^{-1}(I_{cos}/I_{sin})$. The velocity of the S-wave is defined in
terms of  the vector $V=(V_x,V_y)$  in  the usual manner  (Marsh, this
volume);
\[ v(\phi) = \gamma - V_x \cos{2\pi\phi} + V_y \sin{2\pi\phi}  \]
\noindent 
With $\gamma$ the \index{s}{systemic velocity}systemic velocity of the binary. In this prescription, we  need three images  describing the values  of
$I_{avg}$,$I_{cos}$  and   $I_{sin}$  for each   velocity  $(V_x,V_y)$
instead of the conventional  one describing $I_{avg}$. In  other words
the projection from \index{s}{Doppler map}Doppler map to trailed spectrogram is then;
\[
F(v,\phi) = \int \left( I_{avg}(V_x,V_y) + I_{cos}(V_x,V_y) \cos{\phi}
+ I_{sin}(V_x,V_y) \sin{\phi} \right) g(V-v) dV_xdV_y
\]
\noindent 
with $g(V-v)$ describing the local line profile  at a Doppler shift of
$V-v$,  which  is   assumed to   be   a Gaussian   convolved  with the
instrumental resolution.

Keith   Horne implemented  a    similar   extension to  the   \index{s}{filtering}filtered
\index{s}{back projection}back-projection code.  However, cross   talk  among the  three  terms
results in significant artifacts  in the back-projected maps. I choose
to  implement the  extension using a maximum \index{s}{entropy}entropy optimiser in  order to
reconstruct \index{s}{artefact}artefact  free images.  Since the  \index{s}{entropy}entropy is only defined
for positive images, and the cosine and sine  amplitudes can be either
positive  or negative, the problem  was implemented using 5
images to characterise the data. One (positive) average image, and two
(positive) images  for both the cosine  and sine amplitudes. For those
two images one image  reflects positive amplitudes, the other negative
amplitudes. The image  \index{s}{entropy}entropy for each image  is defined in the usual
manner relative to a running default image, and  the data is fitted to
a requested $\chi^2$ while maximising the \index{s}{entropy}entropy of the 5 images.

\subsection{A test reconstruction}

\begin{figure}
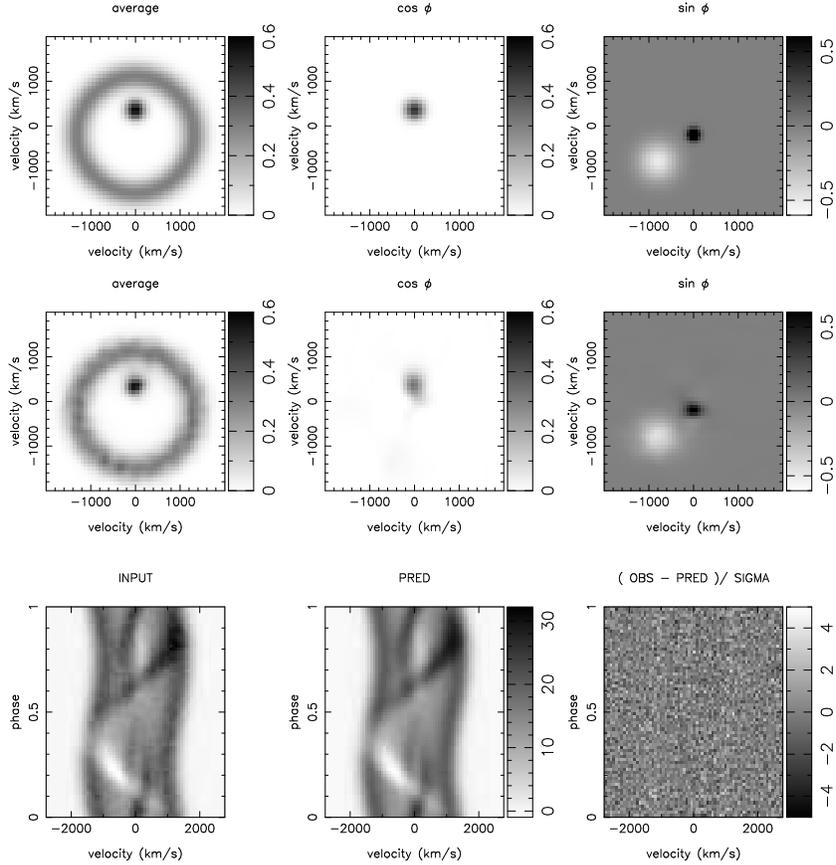

\centerline{\includegraphics*[height=.9\textwidth,angle=-90]{figure13a.ps}}
\centerline{\includegraphics*[height=.9\textwidth,angle=-90]{figure13b.ps}}
\caption[]{A fake data set in order to test the reliability of modulation mapping. Top row plots the input images consisting of emission from a ring and various discrete modulated spots. The data corresponding to those images is plotted in the lower left panel (with \index{s}{random noise}random noise added). Middle row are the reconstructed images, bottom rows compares observed and predicted data, together with a plot of the normalised residuals.}
\label{test}
\end{figure}

In order  to test the method,  and check for possible  artefacts, fake
data was generated  and noise was  added to that data  to see how well
the input images can be reconstructed by the code.  In Figure
\ref{test}, I show an  example of an  input data  set consisting of  a
constant  emission   from  a   disc as well    as  various anisotropic
contributions from 3 spots.   The   large modulation of the    S-waves
corresponding to  the spots  results   in a  trailed  spectrogram that
cannot be fit  using  the conventional \index{s}{Doppler tomography}Doppler tomography   technique.
However, using the  above described extension, the  modulation mapping
code  is able to  fit  the input  data   to a  $\chi^2$ of  1, without
introducing  any spurious features  in any  of  the images, or leaving
systematic residuals to the fit. As expected, the higher the signal to
noise  of   the input  data,  the  more   reliable   and accurate  the
reconstruction  is.  Most  importantly though, there  is no cross-talk
between the  various  images.  A  series of  tests were  performed  to
confirm that the problem is well  constrained and that maximum \index{s}{entropy}entropy
ensures that  no  image structure  is added  unless the  data dictates
it. Two different types of default  were used.  For the average image,
the default is set to a Gaussian  blurred version of the average image
after each iteration, while  for the modulation  images we have  tried
both Gaussian blurring as  well as steering the  images to zero.  Both
converge easily to the maximum \index{s}{entropy}entropy solution. Of course this method
suffers  from the same limitations as  \index{s}{Doppler tomography}Doppler tomography with regards
to errors  in the  assumed  \index{s}{systemic velocity}systemic velocity of the  binary,  limited
image  structure due to  poor signal  to noise, and  \index{s}{artefact}artefacts due  to
limited phase sampling.

\subsection{Application to real data}

\begin{figure}
\centerline{\includegraphics*[width=.35\textwidth,angle=-90]{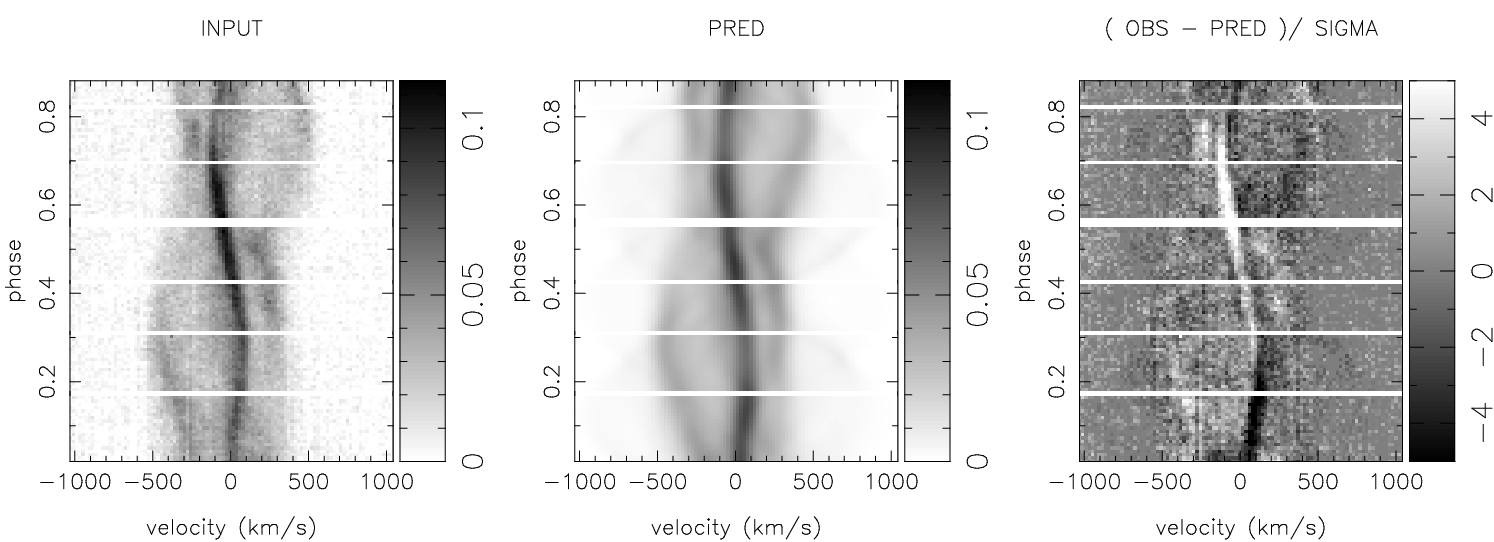}}
\caption[]{The result of \index{s}{Doppler tomography}Doppler tomography of SS Cyg\index{o}{SS Cyg} in \index{s}{outburst}outburst. The observed trailed spectrogram of HeI6678 is on the left, the predicted data of the reconstructed \index{s}{Doppler map}tomogram in the middle panel. The \index{s}{Doppler map}tomogram itself was shown in Figure \ref{others}. Significant residuals (right panel) exist due to the \index{s}{anisotropic radiation}anisotropic emission from the \index{s}{secondary star}secondary star and parts of the disc. }
\label{ss1}
\end{figure}

As    an example of  applying   the method to   real  data, we use the
previously discussed data of SS Cygni\index{o}{SS Cyg} during \index{s}{outburst}outburst (Figure
\ref{others}). Figure    \ref{ss1}  looks  at  the    residuals between
predicted  and observed data  when conventional \index{s}{maximum entropy}MEM \index{s}{Doppler tomography}Doppler tomography
was applied to the  HeI emission line  data.  Although  recovering the
general features in the data, such as the S-wave from the \index{s}{irradiation}irradiated
secondary and the two armed disc  asymmetry, residuals are significant
and the best $\chi^2$ that can  be achieved is  3.2. In particular the
\index{s}{anisotropic radiation}anisotropic emission from the secondary  star  and parts of the  disc
leave  large residuals   due   to their  phase   dependence.   Doppler
tomography tries to reproduce this phase dependence as best as it can,
but  is   fundamentally  unable  to   describe  such  emission sources
adequately.
 
\begin{figure}
\centerline{\includegraphics[width=.7\textwidth,angle=-90]{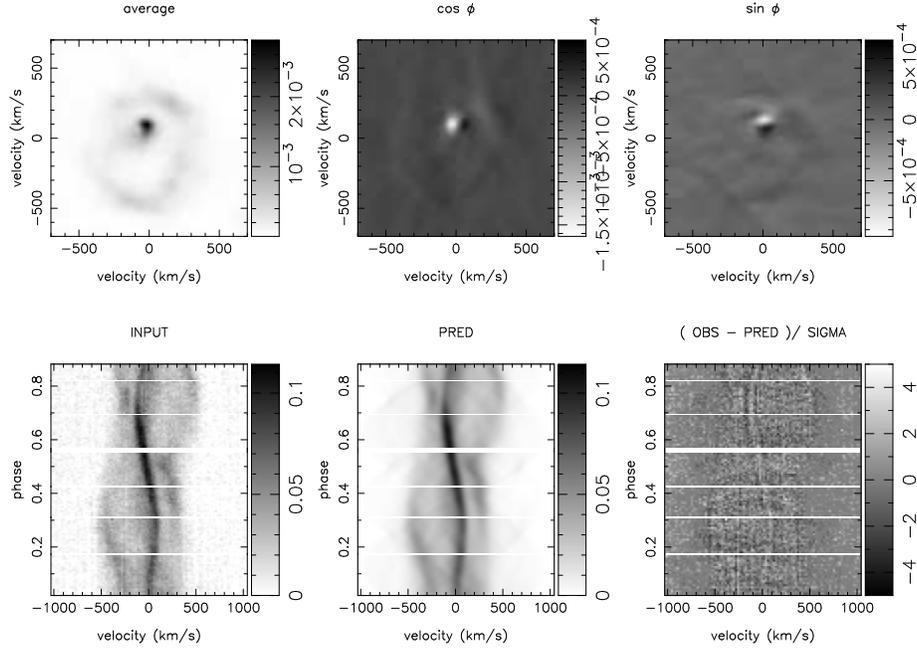}}
\caption[]{The same SS Cygni\index{o}{SS Cyg} data now processed with the modulation mapping code. Top row are the reconstructed images, with the observed and predicted spectrogram in the bottom row.}
\label{ss2}
\end{figure}

If the same data  is  passed to  the  modulation mapping code, a  much
better  fit to  the   data can  be  obtained  (Figure \ref{ss2}).  The
$\chi^2$ value of the fit is 1, as good as  one may expect, and leaves
much reduced residuals between data and fit.   The structure of the
average image is very  similar to that  obtained with the conventional
method, indicating  that the  majority of  the emission is  relatively
unmodulated. In   the  two  modulated  images,  one  can identify   in
particular the emission from the secondary and some  areas of the disc.
 However, the two images tell us not only where the
modulated  emission  is   coming from,   but also  the   phase of  the
modulation.  The \index{s}{secondary star}secondary star emission  shows a left-right asymmetry
in the cosine  image, and a  front-back  asymmetry in the sine  image,
indicating that the emission is indeed beamed away from the \index{s}{Roche equipotential}Roche lobe
surface.   Although one ideally applies  Roche  tomography in order to
model the contribution  of  the secondary, modulation  mapping ensures
that  strong anisotropic S-waves  originating from  the mass donor are
fitted well,   and  that they   do not  leave \index{s}{artefact}artefacts  or  limit the
goodness of fit that can be achieved.

\begin{figure}
\centerline{\includegraphics[height=.9\textwidth,angle=-90]{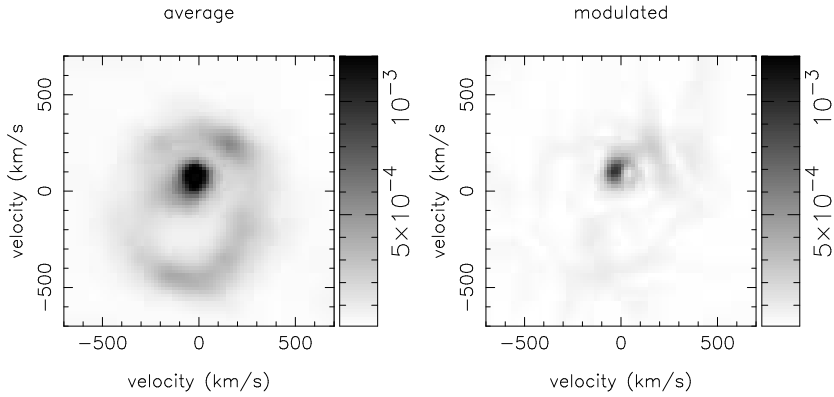}}
\caption[]{A closer look at the derived \index{s}{Doppler map}tomograms with modulation mapping. Left the average image, right the total amplitude in the modulated images. Most prominent is the modulated emission from the \index{s}{secondary star}secondary star, as well as a small contribution from the disc. }
\label{ss3}
\end{figure}

Obvious other  emission sources where  the phasing  and beaming of the
emission can be quantified with this method are  the emission from the
\index{s}{bright spot}bright spot, and emission from anisotropic \index{s}{spiral arms}spiral shocks. In SS Cygni,
the emission from  the disc seems to be  only weakly  modulated (Figure \ref{ss3}), which
may   not   be too      surprising  because   of  its   low    orbital
\index{s}{inclination}inclination $i$.  Shear  broadening,  for  example,  leads  to anisotropic
emission and is and proportional to $\cos^{-1}{i}$. Its effect is thus
considerably reduced for  \index{s}{inclination}inclinations below 60$^{\circ}$ \cite{ss99}.
Work is in  progress to  apply this   method to  the emission of   the
\index{s}{spiral arms}spirals   in  IP Peg  and  other  objects  in order    to quantify the
anisotropy of the emission from the \index{s}{spiral arms}spiral shocks.

Modulation  mapping provides a   straightforward extension to  Doppler
tomography, with a wide range of  applications.  The code as such does
not need to make any assumptions about the  nature of the modulations,
except that the  period of modulation  is the orbital period.  It thus
relaxes one of  the fundamental assumption  of  \index{s}{Doppler tomography}Doppler tomography, in
order to make the  technique more versatile.  Many data  sets already
exist that  could  benefit  from this extension,   but it  will  be of
particular use for the high resolution data sets that will be obtained
with  the  new   large aperture  telescopes.    An even  more  general
prescription can easily be envisaged that attempts to model modulations
on other periods as well.

\section{Conclusions}

\index{s}{Doppler tomography}Doppler tomography of the \index{s}{cataclysmic variable!dwarf nova}dwarf nova IP Peg\index{o}{IP Peg}  has revealed a remarkable
two  armed \index{s}{spiral arms}spiral structure in  its  accretion disc that dominates the
emission from a range  of emission lines  over a large range of radii.
The pattern is observed from the start of the \index{s}{outburst}outburst up to the later
stages of the \index{s}{outburst}outburst maximum, and is fixed in the binary frame.  Its
location  in the \index{s}{Doppler map}tomograms  fits  remarkably well  with the two  armed
\index{s}{spiral arms}spiral shocks that are expected to be generated by the tidal torques
of  the companion  star.  A tidal origin  is also
supported by the  fact  that the structure   is corotating  with  the
binary for  at  least 50   orbital periods,  and   the fact  that  the
asymmetry is not visible during  \index{s}{quiescence}quiescence, when tidal torques on the
much smaller accretion disc are significantly reduced.

In order to appreciate the   relevance of tidally driven spiral  waves
for  the structure  of  accretion  discs   in general, a  more  varied
observational   picture is required, spanning a   range of objects and
source states. Since the detection of \index{s}{spiral arms}spirals in  IP Pegasi\index{o}{IP Peg}, a handful
of  other  systems have   displayed   very similar  disc   structures, 
 but  a systematic picture is still difficult to extract. 
A second  important restriction  of   the current  data sets,  is  the
limited coverage we  have of  each \index{s}{outburst}outburst.  In the CV  sub-class  of
\index{s}{cataclysmic variable!dwarf nova}dwarf novae we have a truly unique opportunity  to track the evolution
of  the  disc in real-time  through   time-lapsed tomography.  What is
needed is a focused campaign that aims to obtain spectroscopy across a
significant fraction of the \index{s}{outburst}outburst cycle.  Such  a data set would be
an extremely  valuable test  bed  for both  disc instability models as
well as  the question of  angular  momentum transport  associated with
density waves. Although the observed \index{s}{spiral arms}spirals appear to fit to detailed
simulation in surprising detail, it is still not clear what the impact
of such prominent \index{s}{spiral arms}spiral arms is on the \index{s}{angular momentum}angular momentum budget of the
disc and how  it relates to the local  shear \index{s}{viscosity}viscosity.  This requires
improved  observations  and    realistic   simulations in    order  to
test quantitatively the impact of such waves (Boffin, this volume).

I also  discussed an extension to  \index{s}{Doppler tomography}Doppler tomography, with the aim of
mapping modulated emission sources in  emission line data.  This is  a
rather common  situation and will  not only  benefit the observational
study of \index{s}{spiral arms}spiral waves, but a  range of other emission sources commonly
observed. I demonstrated that   \index{s}{artefact}artefact free reconstructions can   be
calculated from phase resolved spectroscopy using maximum \index{s}{entropy}entropy regularisation.

%

\end{document}